\documentclass[fp,twocolumn]{jpsj3}
\usepackage{txfonts}
\usepackage{bm}
\usepackage{color}
\usepackage{ulem}

\title{Microscopic Description of Yielding in Glass Based on Persistent Homology}

\author{ Tatsuhiko Shirai$^1$\thanks{tatsuhiko.shirai@aoni.waseda.jp}, and Takenobu Nakamura$^2$}
\inst{$^1$Institute for Solid State Physics, University of Tokyo, 5-1-5 Kashiwa, Chiba 277-8581, Japan \\
$^2$National Institute of Advanced Industrial Science and Technology (AIST),
1-1-1 Umezono, Tsukuba, Ibaraki 305-8568, Japan} %\\

\abst{
Persistent homology (PH) was applied to probe the structural changes of glasses under shear.
PH associates each local atomistic structure in an atomistic configuration to a geometric object, namely, a {\it hole}, and evaluates the robustness of these holes against noise.
We found that the microscopic structures were qualitatively different before and after yielding.
The structures before yielding contained robust holes, the number of which decreased after yielding. 
We also observed that the structures after yielding approached those of quickly quenched glass.
This work demonstrates the crucial role of robust holes in yielding and provides an interpretation based on geometry.
}

\begin{document}
\maketitle

\section{Introduction}
Based on statistical mechanics, the mechanical properties of materials can be explained through geometric objects in an atomistic configuration. 
For example, the compressibility of a simple liquid can be represented by its radial distribution function (RDF)~\cite{hansen1990theory}
and the rigidity of a crystalline solid can be described by its microscopic periodicity 
[i.e., long-range order (LRO)]~\cite{chaikin2000principles}. 
However, understanding the mechanical properties of glasses at the microscopic level remains challenging owing to the lack of an evident geometric object.

Characterizing the atomistic structure of a glass is difficult, even when the configuration data of the glass are available from molecular dynamics (MD) simulations, because the structure observed in the glass is neither the nearest-neighbor pairs nor the LRO associated with an infinite number of atoms.
These structures are referred to as medium-range order (MRO) and are composed of a large but finite number of atoms.
Descriptors for the MRO have previously been developed.
For example, the methods for multi-atom correlations were introduced as an extension of short-range orders, such as the RDF and structure factor~\cite{elliott1983physics,Zallen1983,miracle2003influence,sheng2006atomic}, which were originally used to characterize uniform and isotropic geometries in liquids.
In addition, several descriptors have been used to describe the distortions from crystalline structures, such as Steinhardt--Nelson order parameters~\cite{PhysRevB.28.784}, bond angle, and dihedral angle~\cite{elliott1983physics,Zallen1983}.
Another approach has been proposed based on network topology, which is particularly useful for characterizing the structures of covalent glasses~\cite{LEROUX201070}.

In recent years, one of the authors has found that persistent homology (PH)~\cite{edelsbrunner2010computational} can be used to describe the MRO of glasses~\cite{nakamura2015persistent,hiraoka2016hierarchical}.
The target of PH is ``holes''.
In this approach, the MRO is represented via metrics of the holes referred to as ``birth'' and ``death''.
In contrast to traditional descriptors, PH can quantify the robustness of the shape of the holes against perturbation.
With the aid of this property, we can describe the structural changes occurring during deformation.

The mechanical properties of glasses can be characterized using stress--strain curves. 
The stress--strain curve depends on the cooling rate used to obtain the glass from the liquid state.
Both the shear modulus and yield stress increase with decreasing cooling rate~\cite{ashwin2013cooling}.
The glass created upon cooling the liquid at a sufficiently low rate exhibits yielding~\cite{utz2000atomistic,varnik2004study}.

To date, there have been several studies describing qualitatively different mechanical responses before and after yielding based on embedded atomistic structures~\cite{utz2000atomistic,shi2005strain,albano2005shear,shi2006atomic}.
However, it has also been argued that there is no structural difference before and after yielding in some model glasses~\cite{shi2006atomic,jaiswal2016mechanical}.
This lack of structural difference may be attributable to failure of the method to adequately characterize the atomistic structure, and thus novel methods are still required.

The purpose of this study was to apply PH to probe the structural changes of glasses that occur during shearing.
We herein demonstrate that the microscopic structures are qualitatively different before and after yielding.
We found more robust holes in the structures before yielding than in those after yielding.
Furthermore, we show that the structures after yielding are similar to those of quickly quenched glasses.
We obtained these results by performing molecular dynamics (MD) simulations.

This paper is organized as follows.
Section~\ref{Sec:model} describes the glass model and the technical details of our MD simulations for the cooling and shearing processes.
Section~\ref{Sec:PD} provides a brief introduction to PH and its application to the atomistic configuration of glasses.
Section~\ref{Sec:result} describes the results of our analysis, in which the structural changes occurring during yielding were detected using PH. 
Section~\ref{conclusion} summarizes the paper.

\section{Model and Simulation Details}\label{Sec:model}
\begin{figure}[t]
\centering
\includegraphics[clip,width=0.47\textwidth]{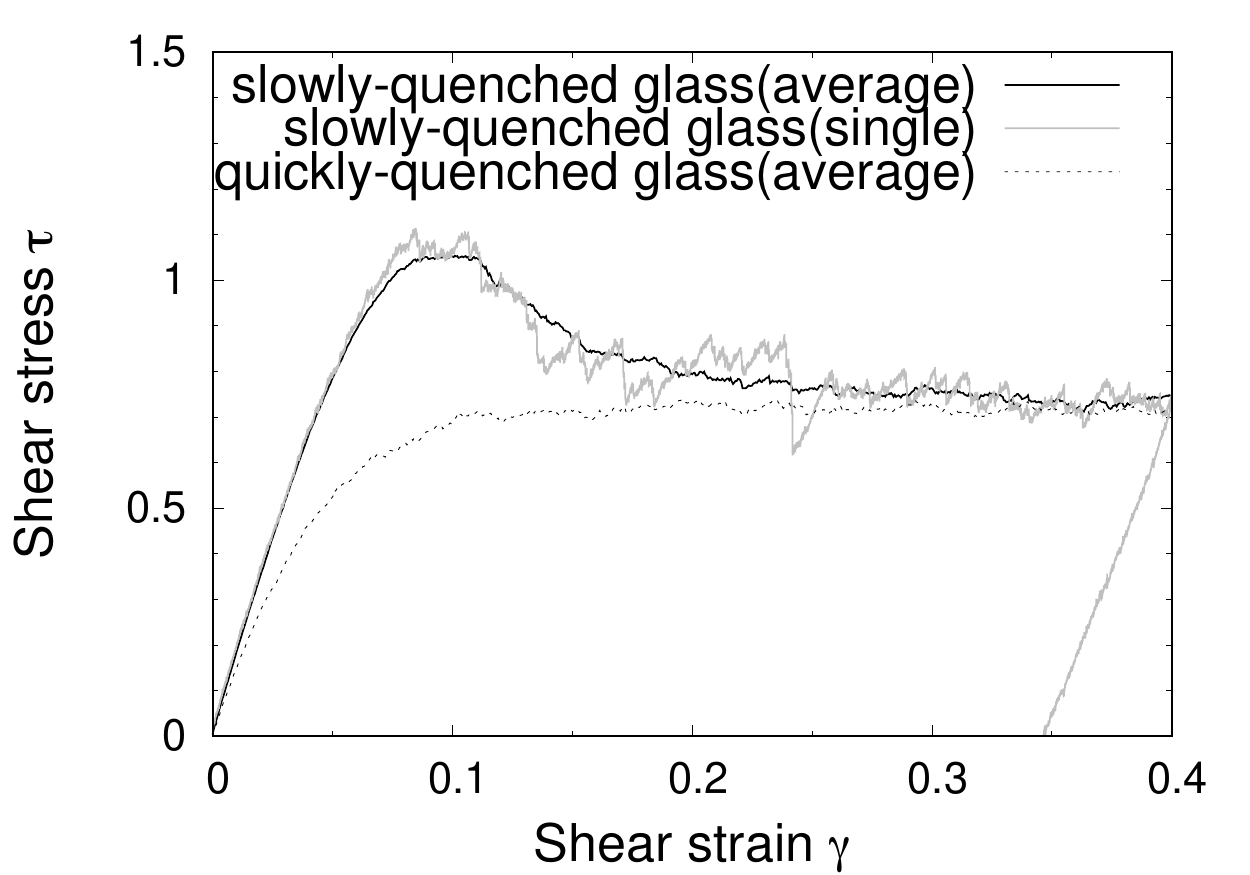}
\caption{
(Color online) Shear stress $\tau$ versus shear strain $\gamma$.
$\tau \propto \gamma$ (linear regime) for the small-shear-strain regime, whereas $\tau \sim$ const. (plateau regime) for the large-shear-strain regime.
The average over 20 independent samples and one particular example for the slowly quenched glass are represented by the black solid line and gray solid line, respectively.
The gray solid line also shows the process of creating a zero-stress state ($\gamma \simeq 0.35$ and $\tau=0$) from the stress state ($\gamma=0.4$ and $\tau \simeq 0.7$).
The dotted line represents the average over 20 independent samples for the quickly quenched glass.
}
\label{stress_strain}
\end{figure}

We consider the glass-forming system referred to as the Kob--Andersen binary Lennard-Jones mixture (KA model)~\cite{kob1995testing}, which consists of two types of particles ($80\%$ large (A) and $20\%$ small (B) particles).
The interaction between $\alpha$ particles and $\beta$ particles ($\alpha, \beta \in \{A, B\}$) is given by the Lennard--Jones potential:
\begin{equation}
U_{\alpha\beta}(r)=4\epsilon_{\alpha\beta}\left[\left(\frac{\sigma_{\alpha\beta}}{r}\right)^{12}-\left(\frac{\sigma_{\alpha\beta}}{r}\right)^6\right],
\end{equation}
where $\epsilon_{AB}=1.5 \epsilon_{AA}$, $\epsilon_{BB}=0.5 \epsilon_{AA}$, $\sigma_{AB}=0.8 \sigma_{AA}$, and $\sigma_{BB}=0.88 \sigma_{AA}$.
A cutoff at $r_c=2.5\sigma_{AA}$ was employed.
The masses of the particles were set as $m_A=m_B$. 
All quantities in this paper are reported in reduced units by setting $\sigma_{AA}$, $\epsilon_{AA}$, and $m_{A}$ to unity.
The number density was set to $\rho=1.2$, and a cubic simulation box was employed with periodic boundary conditions.
We performed MD simulations of this model using the Large-scale Atomic/Molecular Massively Parallel Simulator (LAMMPS)~\cite{plimpton1995fast}.
The time step size was set as $\Delta t=0.005$.
The system contained $N=32000$ particles with a simulation box of size $L=29.876$.
This model is based on metallic glass (${\rm Ni}_{80}{\rm P}_{20}$) and exhibits glass transition at a temperature of $T_c \simeq 0.435$~\cite{kob1995testing}.
Its dynamical properties~\cite{kob1995testing,barrat1999fluctuation,berthier2002shearing} and static properties~\cite{varnik2004study,ashwin2013cooling} have been extensively studied.

We prepared the atomistic configurations of glasses as follows.
First, the system was equilibrated by a $10^6$-step NVT-MD simulation with temperature $T_{\rm i}=1.0$ from an appropriate initial configuration, such as a face-centered cubic lattice or a random configuration.
We used the final configuration from the MD simulation as the liquid configuration.
We obtained 20 statistically independent liquid configurations by repeatedly performing this process.
Then, from each liquid configuration, we created a slowly quenched glass by changing the thermostat temperature with a cooling rate of $\dot{T}=-10^{-3}$.
After reaching almost zero temperature, we performed energy minimization using the conjugate gradient method to convert each system to mechanical equilibrium.
For comparison, a quickly quenched glass was also created by applying energy minimization to the liquid state at $T=1.0$ using the conjugate gradient method.
In total, 20 independent atomistic configurations were prepared for each of the two glasses. 

\begin{figure}[t]
\centering
\includegraphics[clip,width=0.47\textwidth]{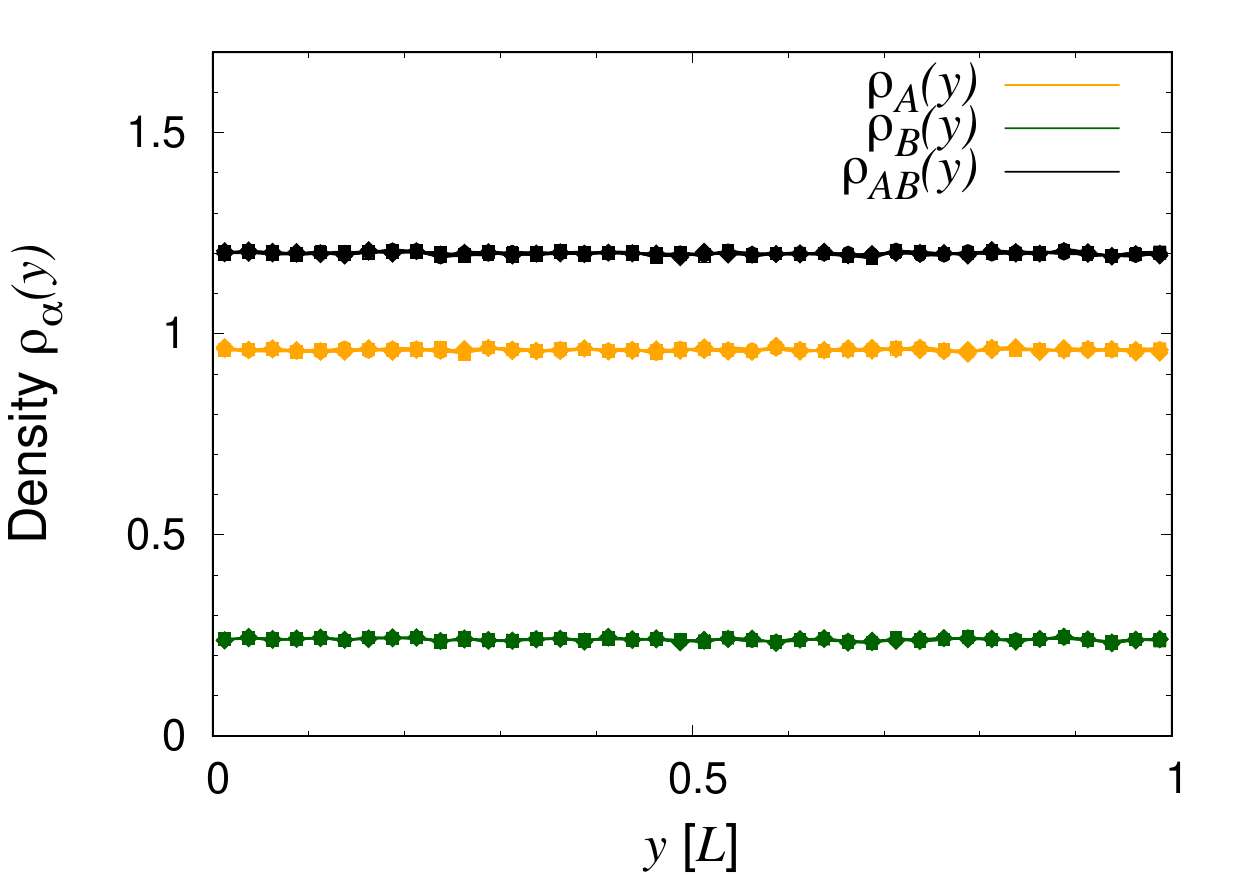}
\caption{
(Color online) Density profiles for $A$ particles ($\rho_A(y)\simeq 0.96$, orange points), $B$ particles ($\rho_B(y)\simeq 0.24$, green points), and $A$ and $B$ particles ($\rho_{AB}(y)\simeq 1.2$, black points) for the sheared states of the slowly quenched glass.
Each data series was averaged over 20 independent samples.
A total of $10$ values of $\gamma$ were selected as $\gamma=0.04 n, n\in\{1,2,\cdots,10\}$, although the density profiles were almost independent of $\gamma$ and mutually indistinguishable.
}
\label{density}
\end{figure}

After quenching, we applied shear using volume-conserving strain deformation with an athermal quasistatic method~\cite{utz2000atomistic,tanguy2006plastic}.
In this method, plastic deformation was imposed on the system by iterating the following deformation step:
First, the position of each particle, $\bm{r}=(x,y,z)$, was transformed into $\bm{r'}=(x-\Delta\gamma \cdot y, y, z)$, where $\Delta \gamma$ is the step size of strain and was set as $\Delta \gamma=10^{-5}$.
Subsequently, energy minimization was performed for the deformed coordinates using the conjugate gradient method.
For each step, the shear stress was calculated and the stress--strain curve was obtained.
In Fig.~\ref{stress_strain}, the stress $\tau$ is the $xy$ component of the stress tensor, which was calculated using a virial, and the strain $\gamma$ is defined as $\gamma =N_{\rm step}\Delta \gamma$, where $N_{\rm step}$ is the number of deformation steps.
For the slowly quenched glass, the stress was linearly dependent on the strain for small strains (linear regime).
After yielding, the stress decreased and finally reached a certain value that was independent of strain (plateau regime).
For the quickly quenched glass, no yielding was observed, and thus it directly reached the plateau regime from the linear regime~\cite{varnik2004study}.
For each stress state, we also calculated the partial and total number density profiles along the $y$ axis and observed uniform profiles in every case (Fig.~\ref{density}).
Therefore, heterogeneous pattern formation did not appear during shear deformation.

\section{Persistent Homology}\label{Sec:PD}
\begin{figure}[t]
\centering
\includegraphics[clip,width=0.4\textwidth]{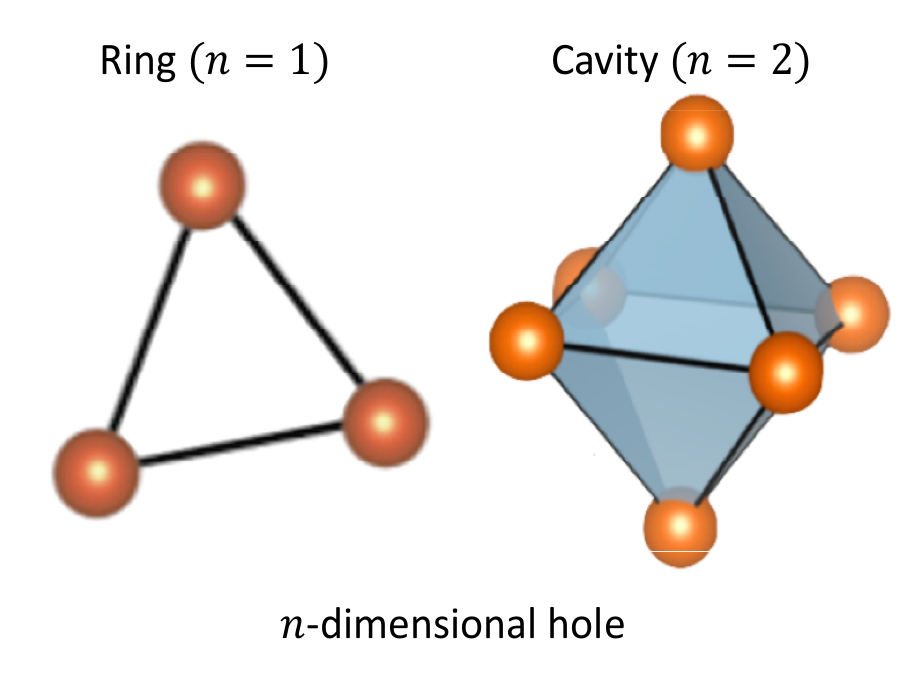}
\caption{
(Color online) Examples of $n$-dimensional holes: ring $(n=1)$ and cavity $(n=2)$.
}
\label{Hole}
\end{figure}
\begin{figure}[t]
\centering
\includegraphics[clip,width=0.5\textwidth]{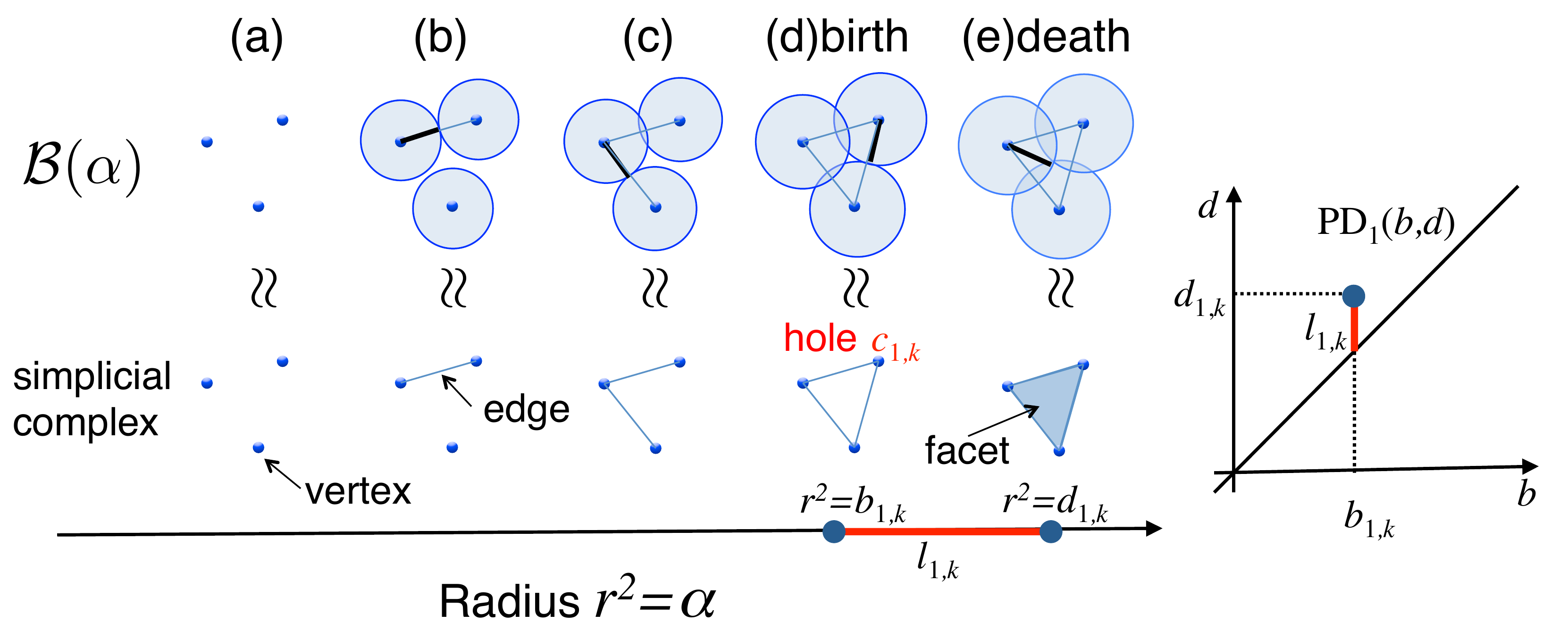}
\caption{
(Color online) PH for the one-dimensional hole in two-dimensional space ($n=1$ and $D=2$).
We set all of the input radii $\{R_i\}$ to zero for simplicity.
The topological features of the union of balls ${\cal B}(\alpha)$ change with the radius of each particle, $r=\sqrt{\alpha}$, from (a) to (e).
The homotopy-equivalent structure of ${\cal B}(\alpha)$ is given by a simplicial complex, which is a set of vertices, edges, and facets.
The homology defines a hole $c_{1,k}$ for the simplicial complex.
PH characterizes the hole using two length scales, $b_{1,k}$ and $d_{1,k}$, where the hole appears [birth; panel (d)] and disappears [death; panel (e)].
The collection of all of the points $(b_{1,k}, d_{1,k})$ provides a PD.
}
\label{PH}
\end{figure}

We provide a systematic method for applying PH to the atomistic configurations of glasses.
PH is a mathematical method for associating a point cloud to the geometric objects of ``holes''. Homology provides a precise mathematical definition of $n$-dimensional holes [ring ($n=1$) and cavity ($n=2$); see Fig.~\ref{Hole}]
and counts the number of holes for a given geometric object referred to as a simplicial complex.
The simplicial complex is composed of vertices, edges, and facets.
In the procedure described below, the PH is used to construct a series of simplicial complexes from a point cloud and provide values of ``birth'', ``death'', and ``life'', which serve as metrics of each hole.

The input of the PH is given by a pair of sets ${\cal A}=(Q, R)$,
where $Q$ is a point cloud with $N$ points and $R=(R_1,...,R_N)$ is a set of input radii.
The point cloud $Q$ corresponds to the set of atomic positions in the glass, namely, $Q=(\bm{r}_1,\bm{r}_2,...,\bm{r}_N)$, where $\bm{r}_i \in {\mathbb{R}}^D$ is the position of the $i$-th atom in $D$-dimensional space (in the present study, $D=3$).
For each atom, we assign a ball, ${\cal B}_i(\alpha)$,
\begin{equation}
{\cal B}_i(\alpha):=\{\bm{r}\in {\mathbb{R}}^D : | \bm{r}-\bm{r}_i|\le r_i(\alpha)\},
\end{equation}
where the radius $r_i(\alpha)$ is given by
\begin{equation}
r_i(\alpha)=\sqrt{R_i^2+\alpha},
\end{equation}
and $|\cdot|$ denotes a usual Euclidean distance.
The parameter $\alpha$ determines the radius of each ball.
We examined the topological features of the union of balls 
${\cal B}(\alpha)=\cup_{i=1}^N{\cal B}_i(\alpha)$ for each value of $\alpha$ (see the upper part of Fig.~\ref{PH}).
Technically, instead of considering ${\cal B}(\alpha)$ itself,
the homotopy-equivalent simplicial complex was considered because it is computationally easy to handle (see the lower part of Fig.~\ref{PH}).
The $k$-th $n$-dimensional hole $c_{n,k}$ appears at $\alpha=b_{n,k}$ [``birth'', see Fig.~\ref{PH}(d)] as $\alpha$ increases from $-{\rm min}_i (R_i)^2$.
This hole persists for a certain interval $\alpha\in [b_{n,k},d_{n,k}]$ with size $l_{n,k}$ [``life'', see Fig.~\ref{PH}(d) and (e)] and finally disappears at $\alpha=d_{n,k}$ [``death'', see Fig.~\ref{PH}(e)].

The PH is visualized using the persistence diagram (PD).
The PD is a two-dimensional scatter plot of holes and the $n$-dimensional PD denoted by ${\rm PD}_n$ is given by a collection of pairs ${\rm PD}_n=\{(b_{n,k},d_{n,k})\}_k$.
The PD has three properties.
First, the inequality $b_{n,k}\le d_{n,k}$ is satisfied because birth occurs before death.
Therefore, the points in ${\rm PD}_n$ only appear above the diagonal.
Second, the PD provides a geometric criterion to select the shape of the holes.
For example, if every input radius is zero (i.e., $R_i=0$), 
an obtuse triangle is not regarded as a hole in ${\rm PD}_1$,
whereas an acute triangle appears as a hole in ${\rm PD}_1$, which persists for a finite interval $[b_{1,k},d_{1,k}]$.
A right triangle is situated in between, and the point in ${\rm PD}_1$ is placed on the diagonal, that is, $b_{1,k}=d_{1,k}$.
Finally, the degree of persistence of a hole is parameterized by ``life'', which represents the robustness to satisfying the criterion against perturbation of the input data.
Suppose that random noise is added to the input data, and the values of $b_{n,k}$ and $d_{n,k}$ change accordingly. 
Then, the hole $c_{n,k}$ with small $l_{n,k}$ may reach the diagonal and violate the criterion, whereas the hole with large $l_{n,k}$ continues to satisfy the criterion.
In this sense, the quantity $l_{n,k}$ represents the robustness of the hole for the criterion in PH.
In the atomistic configuration of glasses, noise in the input data may occur due to external perturbation.
It is natural to expect that there exists a relation between the robustness mathematically introduced in the PH and the robustness of atomistic structures against physical perturbation.
This is the key assumption of our PH analysis when we apply it to a glassy system.

In this paper, we consider a distribution of the PD~\cite{nakamura2015persistent},
\begin{align}
PD_n(b,d)=\sum_{k}\delta(b-b_{n,k})\delta(d-d_{n,k}),
\end{align}
and use $PD_2(b,d)$, that is, the PD of the cavity, as a descriptor of the structures in the three-dimensional glass ($D=3$).
We set all of the input radii to zero (i.e., $R_i=0$).

\section{Structural Changes Before and After Yielding}\label{Sec:result}
\begin{figure}[t]
\centering
\includegraphics[clip,width=0.5\textwidth]{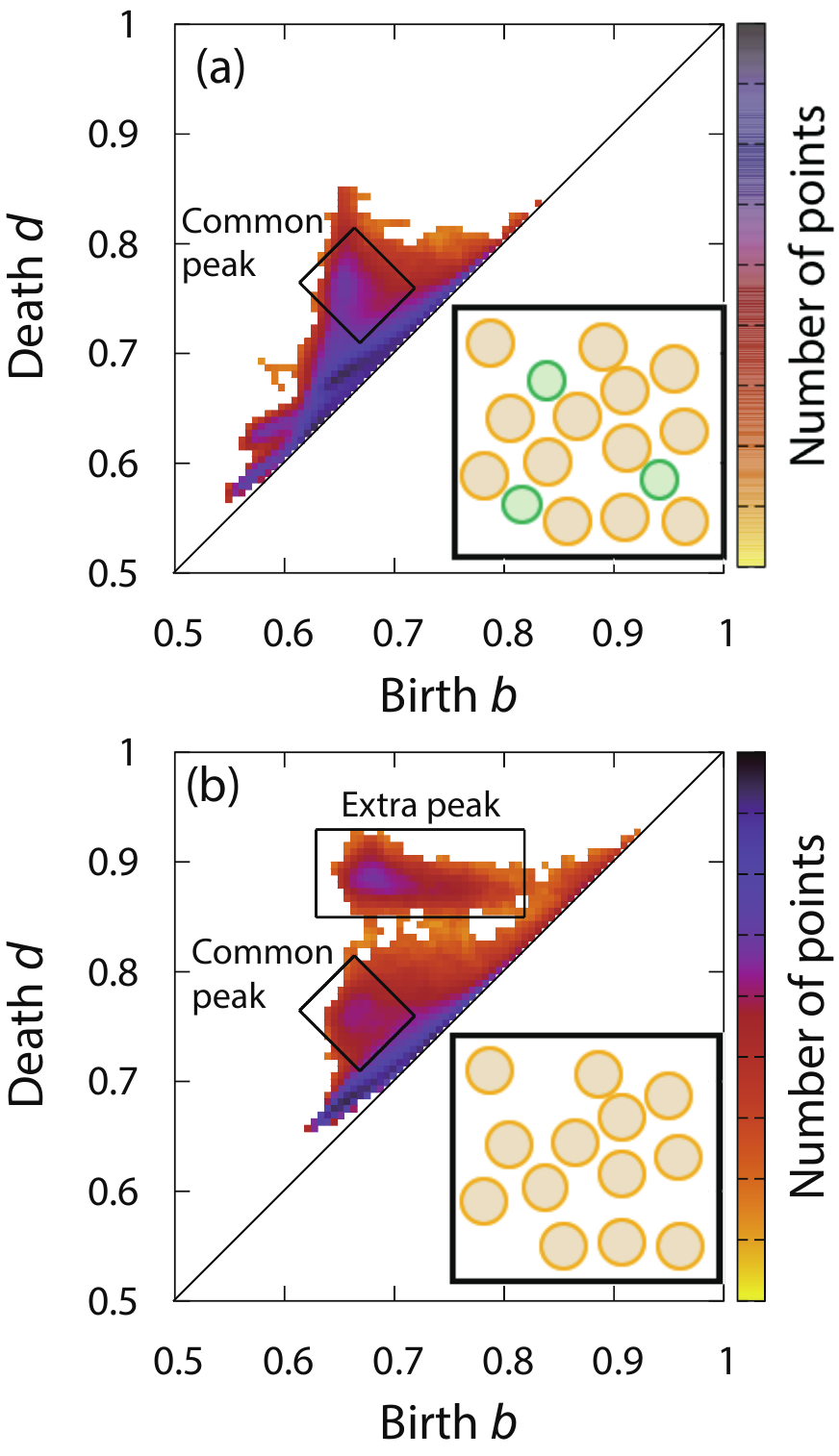}
\caption{
(Color online) $PD_2(b,d)$ for the configurations of $\gamma=0$ of the slowly quenched glass for (a) the configuration of all species and (b) the configuration of only the $A$ species.
The color indicates the number of holes within region $[b,b+\delta) \times [d, d+\delta) (\delta=0.006)$;
the change in color from yellow to black represents an increase in the number of holes.
The peaks far from the diagonal (i.e., those with large $l$) are surrounded by boxes.
The insets show schematic depictions of the atomistic configurations (note that we studied a three-dimensional system).
The $A$ and $B$ particles are represented by orange and green circles, respectively.
}
\label{PD_zero}
\end{figure}

We first provide a PH analysis technique specialized to a several-component mixture system.
Here, we focus on the application of this technique to binary mixture systems, but its extension to several-component systems is straightforward.
For binary mixture systems, we can consider the configuration of only the major species (the A particle in the KA model), in addition to the original configuration of all species.
The former configuration is obtained by removing all of the minor particles from the configuration of all species (see the insets in Fig.~\ref{PD_zero}).
Then, the space occupied by a minor particle---that is, the cage structure around the minor particle---is detected as an additional hole in the PD for the former configuration in addition to the holes originally present in the system.
This cage structure is one of the MROs known as solute-centered polyhedra~\cite{miracle2003influence,sheng2006atomic}.
For example, in $\rm{Ni}_{80}\rm{P}_{20}$ the P-centered tricapped trigonal prism is the most common structural unit~\cite{guan2012structural}.

We applied this technique to the KA model for the zero-stress state with $\gamma=0$ (i.e., the freshly prepared sample) of the slowly quenched glass.
We show the PD for the configuration of all species in Fig.~\ref{PD_zero}(a) and the PD for the configuration of only the $A$ species in Fig.~\ref{PD_zero}(b).
In Fig.~\ref{PD_zero}(b), an extra peak can be observed in the upper region in addition to the peak in the lower region, which is common to Figs.~\ref{PD_zero}(a) and (b).
Consequently, the extra peak and the common peak represent the cage structure surrounded by a $B$ particle and a vacancy between $A$ particles, respectively.
In this manner, we can systematically distinguish two types of holes based on the peak positions in the PD: an extra peak at $(b,d)=(0.68, 0.88)$ and a common peak at $(b,d)=(0.67, 0.76)$~\cite{comment1}.
Hereinafter, we only consider the PD for the configuration of only the $A$ species, as this configuration is sufficient for characterizing the two types of holes.

\begin{figure}[t]
\centering
\includegraphics[clip,width=0.47\textwidth]{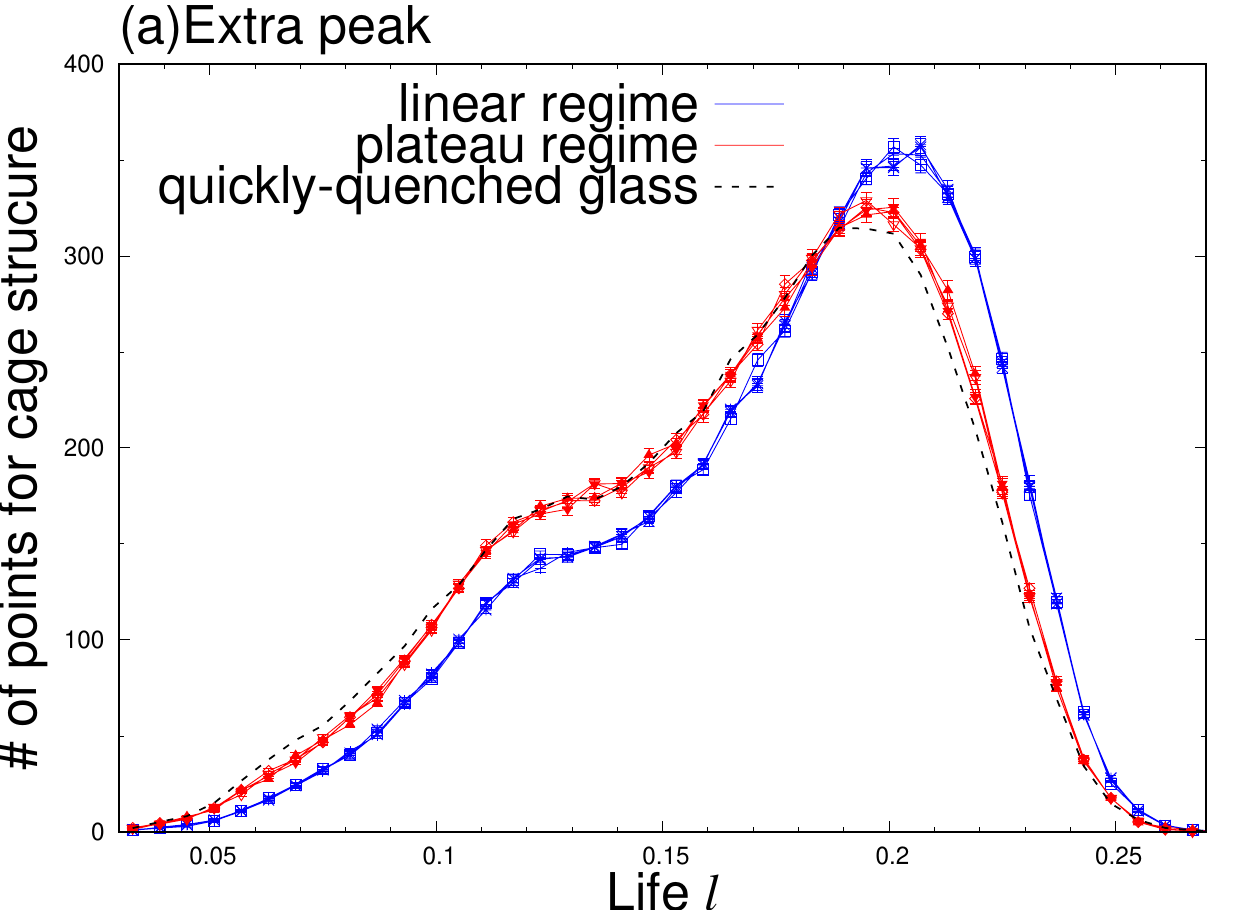}\\
\includegraphics[clip,width=0.47\textwidth]{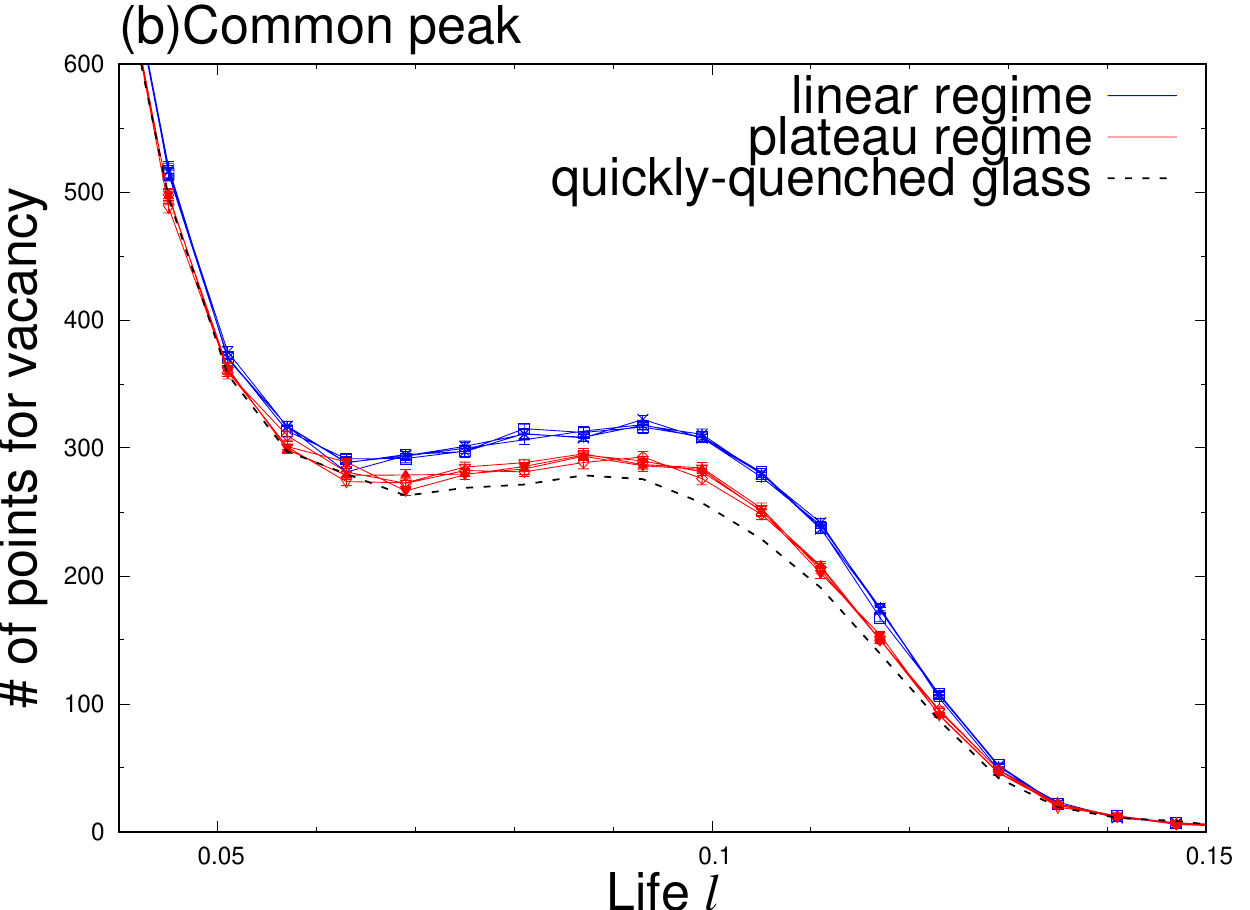}
\caption{
(Color online) Histograms of life $(l=d-b)$ for the configurations of $\gamma$.
Panels (a) and (b) show the extra and common peaks observed in the PD (Fig.~\ref{PD_zero}(b)), respectively.
The linear regime ($\gamma_0 = 0, 0.02, 0.04, 0.06$) and plateau regime ($\gamma_0 = 0.34, 0.36, 0.38, 0.4$) for the slowly quenched glass are represented by blue and red solid lines, respectively.
The histogram for the quickly quenched glass at $\gamma=0$ is plotted as a dotted line for comparison.
}
\label{Life_dist1}
\end{figure}

Based on this interpretation, we investigated the structural changes in the PD during shear deformation.
The stress states exhibited a trivial anisotropic deformation in response to finite shear stress.
To eliminate this contribution, we prepared a zero-stress state for each stress state with strain $\gamma$.
Specifically, we reset the shear stress to zero by decreasing the shear strain in a stepwise manner while minimizing the energy (see the gray solid line in Fig.~\ref{stress_strain}).
In the zero-stress state, the structural change owing to the trivial anisotropic contribution is eliminated.
In this paper, we refer to this state as the zero-stress state with $\gamma_0$~\cite{karmakar2010plasticity}, where $\gamma_0$ is the value prior to decreasing the shear strain~\cite{comment3}.
We calculated the PDs for the zero-stress state with $\gamma_0$.
According to Fig.~\ref{stress_strain}, we regard the zero-stress states with $\gamma_0=0, 0.02, 0.04, 0.06$ as those in the linear regime and the zero-stress states with $\gamma_0=0.34, 0.36, 0.38, 0.4$ as those in the plateau regime.
In the series of PDs, we observed no clear differences within the linear regime or within the plateau regime, 
whereas remarkable differences could be discerned between the two regimes.

Let us consider the extra peak corresponding to the cage structure surrounded by a $B$ particle.
As shown in Fig.~\ref{PD_zero}(b), we introduce a box in the upper region, which is specified by a rectangular region, $(b,d) \in [0.63, 0.82]\times[0.85, 0.93]$.
The histograms of life ($l=d-b$; see Sec.~\ref{Sec:PD}) for the points within this box are plotted in Fig.~\ref{Life_dist1}(a) for the linear regime (blue lines) and plateau regime (red lines).
We found no differences within the error for each regime, indicating no structural changes within the regimes.
However, clear differences were observed between the two regimes, 
namely, the height of the peak around $l=0.2$ decreased, and the number of holes with a smaller $l$ increased.
This result demonstrates that the number of robust holes for the cage structure surrounded by a $B$ particle decreased from the linear regime to the plateau regime.
For comparison, we also present a histogram of life for the quickly quenched glass at $\gamma=0$ (dotted lines).
We found that this was considerably close to that of the zero-stress state of the slowly quenched glass in the plateau regime.
Qualitatively similar behavior was also observed for the common peak [Fig.~\ref{Life_dist1}(b)], which was defined by a box region, $1.38\leq b+d \leq 1.48$ and $0.04 \leq d-b \leq 0.15$ [see Fig.~\ref{PD_zero}(b)].

\begin{figure}[t]
\centering
\includegraphics[clip,width=0.47\textwidth]{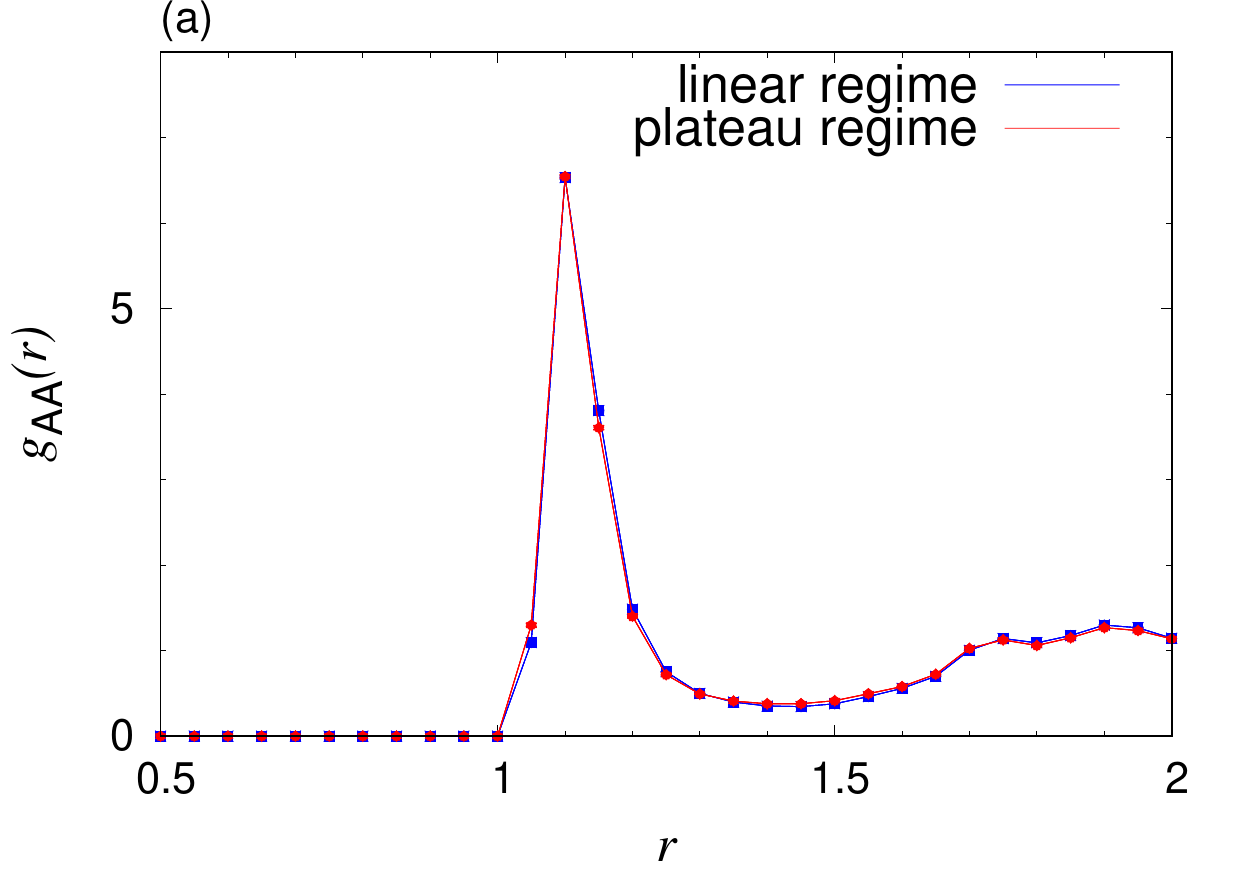}\\\includegraphics[clip,width=0.47\textwidth]{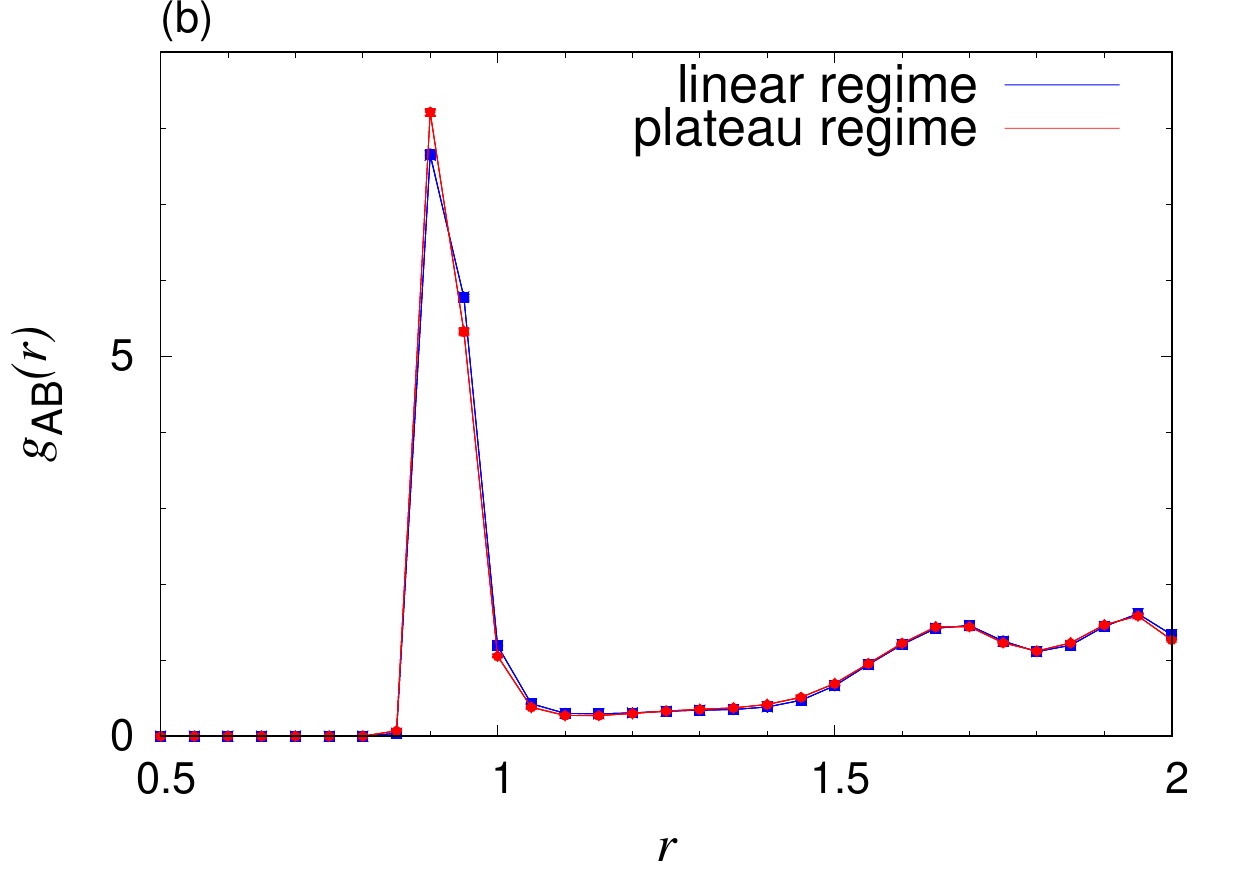}\\\includegraphics[clip,width=0.47\textwidth]{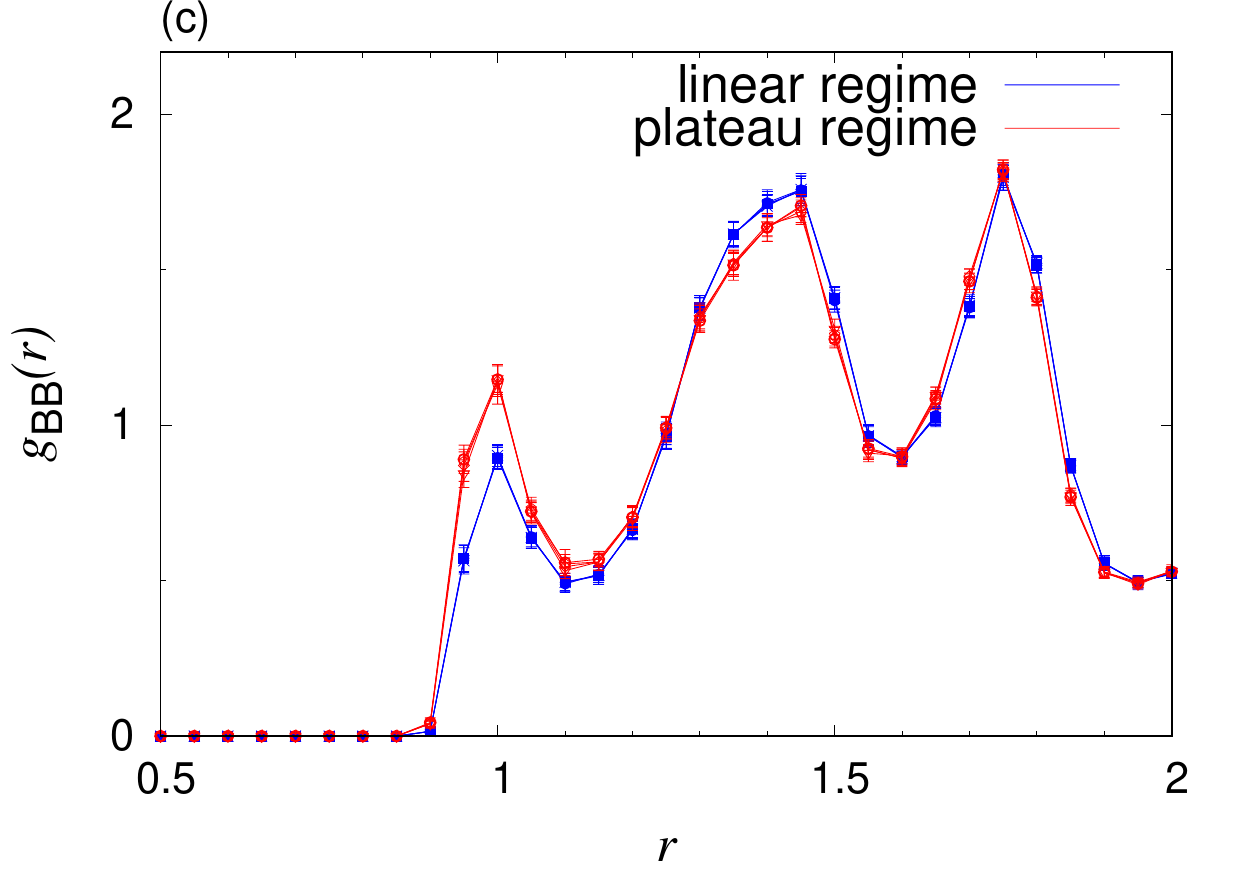}
\caption{
(Color online) Partial RDFs of (a) $A$--$A$, (b) $A$--$B$, and (c) $B$--$B$ particles.
The linear regime ($\gamma_0=0, 0.02, 0.04, 0.06$) and plateau regime ($\gamma_0=0.34, 0.36, 0.38, 0.4$) for the slowly quenched glass are represented by blue and red curves, respectively.
}
\label{gofr}
\end{figure}

For comparison with the PDs, we also calculated the traditional descriptors known as partial RDFs.
As discussed in a previous study~\cite{utz2000atomistic}, structural change of the cage structure surrounded by a $B$ particle was observed in the partial RDF between $B$ particles, $g_{\rm BB}(r)$.
As shown in Fig.~\ref{gofr}(c), the height of the first peak increased from the linear regime to the plateau regime, whereas that of the second peak slightly decreased.
This observation combined with the PH analysis indicates that direct contact between the $B$ particles is suppressed in the linear regime owing to the formation of the cage structure.
In other words, direct contact breaks the robust cage structures during shear deformation.
On the contrary, $g_{\rm AB}(r)$ cannot be used to distinguish the linear and plateau regimes [see Fig.~\ref{gofr}(b)].
Although $g_{\rm AA}$ is a descriptor obtained from only the $A$ species as well as the common peak [Fig.~\ref{Life_dist1}(b)], it also cannot be used to adequately distinguish the linear and plateau regimes [see Fig.~\ref{gofr}(a)].
Thus, the PD can reveal higher-order correlations such as MRO beyond the description of $g_{\rm AA}$.

\begin{figure}[t]
\centering
\includegraphics[clip,width=0.47\textwidth]{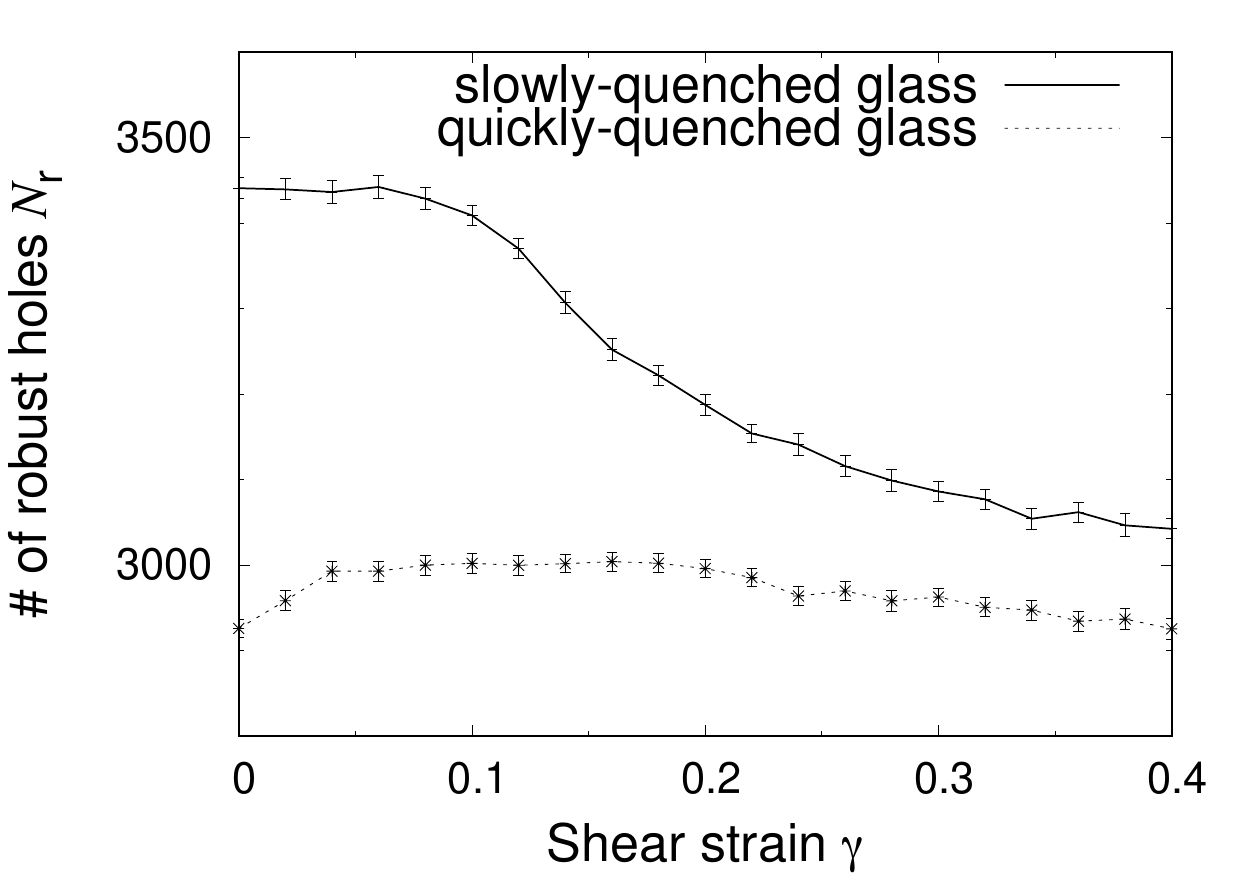}
\caption{
(Color online)
Dependence of the number of robust holes $N_{\rm r}$ on the shear strain $\gamma$.
The solid and dotted lines represent the slowly quenched glass and quickly quenched glass, respectively.
}
\label{Robust}
\end{figure}

As discussed in the preceding text, structural changes upon yielding appear to be explained as a decreased number of robust holes in the PD.
To confirm this scenario quantitatively, we studied the cooling-rate dependence of the structures by focusing on ``robust holes''.
By specifying the region with a long life $l>0.17$ and $(b,d) \in [0.63, 0.82]\times[0.85, 0.93]$ [corresponding to the extra peak in the upper region of Fig.~\ref{PD_zero}(b)],
we counted the number of robust holes, denoted $N_{\rm r}$, which corresponds to the summation of the histogram for $l>0.17$ in Fig.~\ref{Life_dist1}(a).
\cite{comment2}
For the slowly quenched glass (solid line in Fig.~\ref{Robust}), $N_{\rm r}$ remained essentially unchanged until approximately $\gamma=0.1$, at which point the shear stress started to decrease (see Fig.~\ref{stress_strain}).
However, $N_{\rm r}$ decreased during yielding, which clearly demonstrates the close relation between the macroscopic yielding phenomenon and the microscopic structural changes from robust holes to noisy holes.
In contrast, for the quickly quenched glass (dotted line in Fig.~\ref{Robust}), $N_{\rm r}$ was almost independent of $\gamma$ and close to that for the slowly quenched glass at large $\gamma$.
This is referred to as ``rejuvenation'', and it was reported previously based on the estimation of potential energy and the partial RDF~\cite{utz2000atomistic}.
The observation of the monotonic decrease in $N_{\rm r}$ provides strong evidence that the robust holes indeed play a significant role in the mechanical responses of glasses, in spite of the fact that the ``robustness'' is mathematically introduced by the PH.
We propose herein an additional perspective from the geometric point of view; that is, {\it robust} local structures are created during cooling processes and broken by applying a large shear on the system.

\section{Conclusion}\label{conclusion}
This study demonstrates that PH provides a means to probe the structural changes of glasses upon yielding.
With the aid of the PD, we found that the structural changes can be encoded in two types of robust holes in the KA model;
one is the cage structure surrounding a minor particle [the extra peak in Fig.~\ref{PD_zero}(b)], and the other is the vacancy between major particles [common peaks in Figs.~\ref{PD_zero}(a) and (b)].
By using the PH, we quantified that the structure of the slowly quenched glass approached that of the quickly quenched glass after yielding.
We can interpret this result as a decrease in the number of robust holes,  implying that the quantity of ``life'' introduced in the PH plays an important role in yielding.
This interpretation is consistent with the observation that the structures of quickly quenched glass do not exhibit shear-strain dependence.

This work has revealed the intimate relation between robust holes and yielding during the shearing process.
To confirm the validity of this relation, other types of model glasses, such as covalent glass, should also be examined in future work.
Furthermore, the relation between robust holes and existing candidates for plasticity carriers, such as ``soft spots'',~\cite{maloney2006amorphous, manning2011vibrational, patinet2016connecting} should be investigated in future research.

\begin{acknowledgment}
\acknowledgment
The authors would like to thank Tomoaki Niiyama for the helpful discussion.
This work was supported by JSPS KAKENHI (Grant Number 15K13530) and JST-PREST.
\end{acknowledgment}

\bibliographystyle{jpsj}

\begin{thebibliography}{100}

\bibitem{hansen1990theory}
J.-P. Hansen and I.~R. McDonald: {\em Theory of simple liquids} (Elsevier,
  1990).

\bibitem{chaikin2000principles}
P.~M. Chaikin and T.~C. Lubensky: {\em Principles of condensed matter physics}
  (Cambridge university press, 2000).

\bibitem{elliott1983physics}
S.~R. Elliott: {\em Physics of amorphous materials} (Longman London; New York,
  1983).

\bibitem{Zallen1983}
R.~Zallen: {\em The Physics of Amorphous Solids} (Wiley-VCH Verlag GmbH, 1983).

\bibitem{miracle2003influence}
D.~Miracle, W.~Sanders, and O.~Senkov: Philos. Mag. {\bfseries 83} (2003) 2409.

\bibitem{sheng2006atomic}
H.~Sheng, W.~Luo, F.~Alamgir, J.~Bai, and E.~Ma: Nature {\bfseries 439} (2006)
  419.

\bibitem{PhysRevB.28.784}
P.~J. Steinhardt, D.~R. Nelson, and M.~Ronchetti: Phys. Rev. B {\bfseries 28}
  (1983) 784.

\bibitem{LEROUX201070}
S.~L. Roux and P.~Jund: Comput. Mater. Sci. {\bfseries 49} (2010) 70.

\bibitem{edelsbrunner2010computational}
H.~Edelsbrunner and J.~Harer: {\em Computational topology: an introduction}
  (American Mathematical Soc., 2010).

\bibitem{nakamura2015persistent}
T.~Nakamura, Y.~Hiraoka, A.~Hirata, E.~G. Escolar, and Y.~Nishiura:
  Nanotechnology {\bfseries 26} (2015) 304001.

\bibitem{hiraoka2016hierarchical}
Y.~Hiraoka, T.~Nakamura, A.~Hirata, E.~G. Escolar, K.~Matsue, and Y.~Nishiura:
  PNAS {\bfseries 113} (2016) 7035.

\bibitem{ashwin2013cooling}
J.~Ashwin, E.~Bouchbinder, and I.~Procaccia: Phys. Rev. E {\bfseries 87} (2013)
  042310.

\bibitem{utz2000atomistic}
M.~Utz, P.~G. Debenedetti, and F.~H. Stillinger: Phys. Rev. Lett {\bfseries 84}
  (2000) 1471.

\bibitem{varnik2004study}
F.~Varnik, L.~Bocquet, and J.-L. Barrat: J. Chem. Phys {\bfseries 120} (2004)
  2788.

\bibitem{shi2005strain}
Y.~Shi and M.~L. Falk: Phys. Rev. Lett {\bfseries 95} (2005) 095502.

\bibitem{albano2005shear}
F.~Albano and M.~L. Falk: J. Chem. Phys {\bfseries 122} (2005) 154508.

\bibitem{shi2006atomic}
Y.~Shi and M.~L. Falk: Phys. Rev. B {\bfseries 73} (2006) 214201.

\bibitem{jaiswal2016mechanical}
P.~K. Jaiswal, I.~Procaccia, C.~Rainone, and M.~Singh: Phys. Rev. Lett.
  {\bfseries 116} (2016) 085501.

\bibitem{kob1995testing}
W.~Kob and H.~C. Andersen: Phys. Rev. E {\bfseries 51} (1995) 4626.

\bibitem{plimpton1995fast}
S. Plimpton: Journal of computational physics {\bfseries 117} (1995) 1.

\bibitem{barrat1999fluctuation}
J.-L. Barrat and W.~Kob: EPL {\bfseries 46} (1999) 637.

\bibitem{berthier2002shearing}
L.~Berthier and J.-L. Barrat: Phys. Rev. Lett. {\bfseries 89} (2002) 095702.

\bibitem{tanguy2006plastic}
A.~Tanguy, F.~Leonforte, and J.-L. Barrat: Eur. Phys. J. E {\bfseries 20}
  (2006) 355.

\bibitem{guan2012structural}
P.~F. Guan, T.~Fujita, A.~Hirata, Y.~H. Liu, and M.~W. Chen: Phys. Rev. Lett.
  {\bfseries 108} (2012) 175501.

\bibitem{comment1} To be precise, another peak exists aside from the diagonal.
This peak is extremely close to the diagonal and also observed in the PD of the liquid configuration~\cite{nakamura2015persistent}.
Therefore, we regard it as a characteristic of the short-range order observed in liquids and neglect it.

\bibitem{karmakar2010plasticity}
S.~Karmakar, E.~Lerner, and I.~Procaccia: Phys. Rev. E {\bfseries 82} (2010) 026104.

\bibitem{comment3} It is noted that in Ref.~\cite{karmakar2010plasticity} $\gamma_0$ is defined as the value of the stress strain {\it after} reaching the zero shear stress, so the definition is different from ours.

\bibitem{comment2} We focus only on the peak in the upper region of Fig~\ref{PD_zero}(b), but the qualitatively same behavior is observed for the peak in the lower region of Fig~\ref{PD_zero}(b)(not shown).

\bibitem{manning2011vibrational}
M.~L. Manning and A.~J. Liu: Phys. Rev. Lett. {\bfseries 107} (2011) 108302.

\bibitem{maloney2006amorphous}
C.~E. Maloney and A.~Lema\^{\i}tre: Phys. Rev. E {\bfseries 74} (2006) 016118.

\bibitem{patinet2016connecting}
S.~Patinet, D.~Vandembroucq, and M.~L. Falk: Phys. Rev. Lett. {\bfseries 117}
  (2016) 045501.
\end{thebibliography}

\end{document}